\documentclass[a4paper,10pt]{article}
\usepackage{graphicx}
\usepackage{comment}
\usepackage{amssymb}
\usepackage{amsmath}
\usepackage{nicefrac}
\usepackage{graphicx}
\usepackage{dcolumn}
\usepackage{bm}
\usepackage{comment}
\usepackage{multirow}

\begin{document}

\huge

\begin{center}
Optimized recursion relation for the computation of partition functions in the superconfiguration approach
\end{center}

\vspace{0.5cm}

\large

\begin{center}
Jean-Christophe Pain$^{a,}$\footnote{jean-christophe.pain@cea.fr}, Franck Gilleron$^a$ and Brian G. Wilson$^b$
\end{center}

\normalsize

\begin{center}
$^a$CEA, DAM, DIF, F-91297 Arpajon, France\\
$^b$Lawrence Livermore National Laboratory, P. O. Box 808, Livermore, California 94550, USA
\end{center}

\vspace{0.5cm}

\begin{abstract}
Partition functions of a canonical ensemble of non-interacting bound electrons are a key ingredient of the super-transition-array approach to the computation of radiative opacity. A few years ago, we published a robust and stable recursion relation for the calculation of such partition functions. In this paper, we propose an optimization of the latter method and explain how to implement it in practice. The formalism relies on the evaluation of elementary symmetric polynomials, which opens the way to further improvements.
\end{abstract}

\newcommand{\bin}[2]{\left(\begin{array}{c}\!#1\!\\\!#2\!\end{array}\right)}

\section{Introduction}\label{sec1}

The super-transition-array (STA) technique is a powerful tool to compute the emission and opacity of intermediate to high-Z plasmas \cite{BARSHALOM89,BLENSKI00,PORCHEROT11,PAIN15,KRIEF15,KRIEF18,WILSON15,HAZAK12,KURZWEIL13,KURZWEIL16} where detailed-configuration-accounting methods would have a prohibitive numerical cost. It was successfully applied to the interpretation of spectroscopy experiments \cite{POIRIER19,DOZIERES19,NAGAYAMA19}. The STA formalism, which is also applicable to non-local-thermodynamic-equilibrium plasmas \cite{BAUCHE04}, requires the calculation of independent-electron partition functions under the constraint that a group of subshells (referred to as a supershell) has an integer number of electrons. Bar-Shalom \emph{et al.} proposed efficient recursion relations (referred to as BS in the following) for generating these partition functions \cite{BARSHALOM89}. Unfortunately, in the case of high-degeneracy supershells and/or at low temperature, such relations suffer from numerical instability due to precision cancellations arising from sums of large terms of alternating sign \cite{WILSON99}. The instability occurs in particular when the thermal energy becomes significantly smaller than the energy spread in a supershell (i.e. the dispersion of the energies of the subshells). In 2004, two of us (Gilleron and Pain, hereafter GP) proposed a stable and robust algorithm (see Eq. (25) in Sec. III of Ref. \cite{GILLERON04}), relying on the computation of partition functions by nested recursion, building up supershells one subshell by one subshell, at each stage from ``parent'' supershells (of one less subshell) with smaller numbers of electrons. All the terms entering the sums of this recursion are positive definite, and cancellation effects are therefore avoided. Three years later, in a paper entitled ``Further stable methods for the calculation of partition functions in the superconfiguration approach'' \cite{WILSON07} (hereafter WGP), we proposed two improvements of the initial method:

\begin{itemize}

\item The first one consists in generalizing the recursion relation to holes, when a supershell is more than half-filled with electrons.

\item The second one consists in precomputing some partition functions and storing the results. It relies on the successive use of generating functions with reduced degeneracies.

\end{itemize}

The latter improvement, however, was briefly explained (see the second part of Sec. III in Ref. \cite{WILSON07}, from Eqs. (17) to (23)), and several researchers wrote to us because they did not manage to understand how to proceed. This is mainly due to the fact that Eq. (23) of Ref. \cite{WILSON07} may be misleading. In the present work, we would like to clarify that point. We also found, which was not mentioned in the previous papers \cite{GILLERON04} and \cite{WILSON07}, that the functions $h$, cornerstones of the algorithm, are elementary symmetric polynomials. Many efficient algorithms for the evaluation of the latter quantities are described in the literature.  The WGP method is explained in section \ref{sec2} and the corresponding algorithm is summarized in section \ref{sec3}. The case where the degeneracy of a subshell (or more) is reduced by more than one, is discussed in section \ref{sec4}. 

\section{Efficient computation of canonical partition functions}\label{sec2}

Let us consider a system (supershell) of $Q$ electrons in $N$ subshells (or spin orbitals) with degeneracies $\vec{g}=\{g_1,g_2,\cdots,g_N\}$. The partition function of such a system reads

\begin{equation}
U_{Q,N}\left[\vec{g}\right]=\underbrace{\sum_{p_1=0}^{g_1}\sum_{p_2=0}^{g_2}\cdots\sum_{p_N=0}^{g_N}}_{\sum_{s=1}^Np_s=Q}\prod_{s=1}^N\bin{g_s}{p_s}X_s^{p_s},
\end{equation}

\noindent where $X_s=e^{-\beta(\epsilon_s-\mu)}$, $\beta=1/(k_BT)$, $\mu$ is the chemical potential and $\epsilon_s$ the energy of orbital $s$. The average population of subshell $\alpha$ can be expressed as

\begin{equation}\label{lin}
\langle n_{\alpha}\rangle=\frac{g_{\alpha}}{1+\frac{U_{Q,N}\left[\vec{g}-\vec{1}_{\alpha}\right]}{X_{\alpha}U_{Q-1,N}\left[\vec{g}-\vec{1}_{\alpha}\right]}},
\end{equation}

\noindent where notation $\vec{g}-\vec{1}_{\alpha}$ means that the degeneracy of subshell $\alpha$ is reduced by one (therefore, $\vec{g}-\vec{1}_{\alpha}-\vec{1}_{\beta}$ means that the degeneracies of subshells $\alpha$ and $\beta$ are both reduced by one, \emph{etc.}). In the same way, the quadratic mean reads

\begin{equation}\label{quad}
\langle n_{\alpha}\left(g_{\beta}-n_{\beta}\right)\rangle=\langle n_{\alpha}\rangle g_{\beta}\frac{U_{Q-1,N}\left[\vec{g}-\vec{1}_{\alpha}-\vec{1}_{\beta}\right]}{U_{Q-1,N}\left[\vec{g}-\vec{1}_{\alpha}\right]}.
\end{equation}

\noindent According to Eq. (17) of Ref. \cite{WILSON07}, the corresponding canonical partition function is equal to

\begin{equation}
U_{Q,N}\left[\vec{g}-\vec{1}_{\alpha}\right]=\frac{1}{Q!}\lim_{z\rightarrow 0}\frac{\partial^Q}{\partial z^Q}\left\{F\left[\vec{g}-\vec{1},z\right]H^{(\alpha)}(z)\right\},
\end{equation}

\noindent where $\vec{1}=\{1,\cdots,1\}$, $\vec{1}_{\alpha}=\{\delta_{1,\alpha},\delta_{2,\alpha},\cdots,\delta_{N,\alpha}\}$,

\begin{equation}
F\left[\vec{g}-\vec{1},z\right]=\prod_{s=1}^N\left(1+zX_s\right)^{g_s-1}
\end{equation}

\noindent and

\begin{equation}
H^{(\alpha)}(z)=\prod_{s=1,s\ne\alpha}^N\left(1+zX_s\right).
\end{equation}

\noindent After algebraic manipulation \cite{WILSON07}, one gets

\begin{equation}
U_{Q,N}\left[\vec{g}-\vec{1}_{\alpha}\right]=\sum_{j=0}^{\min(Q,N-1)}h_j^{(\alpha)}~U_{Q-j,N}\left[\vec{g}-\vec{1}\right],
\end{equation}

\noindent with

\begin{equation}\label{hj}
h_j^{(\alpha)}=\lim_{z\rightarrow 0}\frac{1}{j!}\frac{\partial^j}{\partial z^j}H^{(\alpha)}.
\end{equation}

\noindent In the same way, we have:

\begin{equation}
U_{Q,N}\left[\vec{g}-\vec{1}_{\alpha}-\vec{1}_{\beta}\right]=\sum_{j=0}^{\min(Q,N-2)}h_j^{(\alpha,\beta)}~U_{Q-j,N}\left[\vec{g}-\vec{1}\right]
\end{equation} 

\noindent and subsequently

\begin{equation}
U_{Q,N}\left[\vec{g}-\vec{1}_{\alpha}-\vec{1}_{\beta}-\vec{1}_{\gamma}\right]=\sum_{j=0}^{\min(Q,N-3)}h_j^{(\alpha,\beta,\gamma)}~U_{Q-j,N}\left[\vec{g}-\vec{1}\right].
\end{equation} 

\noindent The efficiency comes from precomputing $U_{Q-j}\left[\vec{g}-\vec{1}\right]$ by GP's method (or the recursion over holes, as in WGP) and storing the results. The latter method relies on the relation

\begin{equation}\label{gp}
U_{Q,N}=\sum_{i=0}^{\min(Q,g_N)}\bin{g_N}{i}~X_N^{i}~U_{Q-i,N-1},
\end{equation}

\noindent initialized with $U_{Q,0}=\delta_{Q,0}$, $\delta$ being Kronecker's symbol. The $h_j^{(\alpha)}$ and $h_j^{(\alpha,\beta)}$ are trivially computed on demand. However, it is stated in Ref. \cite{WILSON07} that, for example, $h_j^{(\alpha)}$ is obtained by an $N-1$ (number of subshells in the supershell minus one) step array update

\begin{equation}\label{wrong}
h_j^{(\alpha)}\leftarrow h_j^{(\alpha)}+X_mh_{j-1}^{(\alpha)},
\end{equation}

\noindent with $m=0,\cdots, N$, $m\ne\alpha$. Such an expression was misunderstood by several readers, limiting its practical use. Indeed, Eq. (\ref{wrong}) may suggest that

\begin{equation}
h_j^{(\alpha)}=\sum_{m=1,m\ne\alpha}^Nh_{j-1}^{(\alpha,m)}X_m,\nonumber
\end{equation}

\noindent which is not true. In fact, one has

\begin{eqnarray}
\frac{\partial^j}{\partial z^j}H^{(\alpha)}(z)&=&\frac{\partial^{j-1}}{\partial z^{j-1}}\frac{\partial}{\partial z}\prod_{s=1,s\ne\alpha}^N\left(1+zX_s\right)\nonumber\\
&=&\frac{\partial^{j-1}}{\partial z^{j-1}}\left\{\sum_{m=1,m\ne\alpha}^N\left[\prod_{s=1,s\ne\alpha,s\ne m}^N\left(1+zX_s\right)\right]\frac{\partial}{\partial z}\left(1+zX_m\right)\right\}.
\end{eqnarray}

\noindent Taking the limit $z\rightarrow 0$ and using the notation

\begin{equation}
h_{k}^{(\alpha,m)}=\frac{1}{k!}\frac{\partial^k}{\partial z^k}\left.\prod_{s=1,s\ne\alpha,s\ne m}^N\left(1+zX_s\right)\right|_{z=0},
\end{equation}

\noindent we get

\begin{equation}
j!\times h_j^{(\alpha)}=\sum_{m=1,m\ne\alpha}^NX_m\frac{\partial^{j-1}}{\partial z^{j-1}}\left.\prod_{s=1,s\ne\alpha,s\ne m}^N\left(1+zX_s\right)\right|_{z=0},
\end{equation}

\noindent which yields

\begin{equation}\label{resu}
h_j^{(\alpha)}=\frac{1}{j}\sum_{m=1,m\ne\alpha}^Nh_{j-1}^{(\alpha,m)}X_m.
\end{equation}

\noindent As an example, let us write down the expansion of the functions $H^{(\alpha)}$, $h_j^{(\alpha)}$ and $h_j^{(\alpha,m)}$ for $\alpha=5$ with $N$=5 shells in the supershell:

\begin{eqnarray}
H^{(5)}(z)&=&(1+zX_1)(1+zX_2)(1+zX_3)(1+zX_4)\nonumber\\
&=&1+z(X_1+X_2+X_3+X_4)\nonumber\\
& &+z^2(X_1X_2+X_1X_3+X_1X_4+X_2X_3+X_2X_4+X_3X_4)\nonumber\\
& &+z^3(X_1X_2X_3+X_1X_2X_4+X_1X_3X_4+X_2X_3X_4)\nonumber\\
& &+z^4(X_1X_2X_3X_4).
\end{eqnarray}

\noindent We get

\begin{eqnarray}
h_0^{(5)}&=&1\nonumber\\
h_1^{(5)}&=&X_1+X_2+X_3+X_4\nonumber\\
h_2^{(5)}&=&X_1X_2+X_1X_3+X_1X_4+X_2X_3+X_2X_4+X_3X_4\nonumber\\
h_3^{(5)}&=&X_1X_2X_3+X_1X_2X_4+X_1X_3X_4+X_2X_3X_4\nonumber\\
h_4^{(5)}&=&X_1X_2X_3X_4.
\end{eqnarray}

\noindent The procedure consists in precomputing $U_{Q-j,N}\left[\vec{g}-\vec{1}\right]$ by the double recursion method proposed in Ref. \cite{GILLERON04}, combined with the recursion over holes published in Ref. \cite{WILSON07}, and storing the results. The coefficients $h_j^{(\alpha\beta\gamma\cdots)}$ can be determined ``on demand'' by the relation (\ref{resu}).

More precisely, let us consider that we have already omitted the orbitals $\alpha,\beta, ...$ and that we are working with a set of $n$ orbitals with $n<N$. The coefficients $h$ (written $\sigma$ in the following to simplify the notations) are in fact the elementary symmetric polynomials 

\begin{equation}
\sigma_{j,n}=\sum_{1\leq \pi_1<\pi_2<\cdots<\pi_j\leq n}X_{\pi_1}X_{\pi_2}\cdots X_{\pi_j}
\end{equation}

\noindent and obey the following recursion relation (see \ref{appa}):

\begin{equation}\label{recuelem}
\sigma_{j,n}=\sigma_{j,n-1}+X_n\sigma_{j-1,n-1},
\end{equation}

\noindent initialized with $\sigma_{j,0}=\delta_{j,0}$, and which is precisely relation (\ref{wrong}), or Eq. (23) of Ref. \cite{WILSON07}. Note that Eq. (\ref{recuelem}) is the analogue of the GP recursion relation applied for $g_i$=1 $\forall i$. It is also possible to consider the recurrence in the other way, \emph{i.e.} to express the coefficients $\sigma$ for $n$ subshells in terms of the coefficients $\sigma$ for $n+1$ subshells (see \ref{appb}).

\section{Algorithm}\label{sec3}

Finally, the algorithm can be summarized as follows:

\begin{itemize}

\item Express, using Eqs. (\ref{lin}) and (\ref{quad}), population averages in terms of modified partition functions of the kind $U_{Q,N}\left[\vec{g}-\vec{1}_{s_1}-\vec{1}_{s_2}-\cdots-\vec{1}_{s_k}\right]$, where $\vec{g}-\vec{1}_{s_1}-\vec{1}_{s_2}-\cdots-\vec{1}_{s_k}$ means that the degeneracies of subshells $s_1$, $s_2$, ..., and $s_k$ are reduced by one. Consider the reduced set of $n=N-k$ subshells where subshells $s_1$, $s_2$, ..., $s_k$ have been omitted. In order to calculate the latter partition functions using the recursion relation

\begin{equation}
U_{Q,N}\left[\vec{g}-\vec{1}_{s_1}-\vec{1}_{s_2}-\cdots-\vec{1}_{s_k}\right]=\sum_{j=0}^{\min(Q,N-1)}\sigma_{j,N-k}~U_{Q-j,N}\left[\vec{g}-\vec{1}\right];
\end{equation}

\item Compute the quantity $\sigma_{j,N-k}$ with the recursion relation

\begin{equation}\label{reln}
\sigma_{j,n}=\sigma_{j,n-1}+X_n\sigma_{j-1,n-1},
\end{equation}

\noindent initialized by $\sigma_{j,0}=\delta_{j,0}$. Since $\sigma_{j,n}$ are nothing else than the elementary symmetric polynomials, they can be obtained by different ways, for instance resorting to formulas (\ref{sym}), (\ref{bell}) or (\ref{det}). Equation (\ref{reln}) can be viewed as a particular case of the GP relation with a beforehand reduced set of subshells with all the degeneracies fixed at the value one. In that framework, the quantities $\sigma_{j,n}$ are obtained by an array update performed using in-place memory, which means that separate input and output arrays are not needed in their generation. More precisely (we omit deliberately the second index $n$ to emphasize the fact that we handle a single vector):

\vspace{0.5cm}

\begin{itemize}

\item Intitialization: $a_i=\delta_{i,0}$, $i$=0 to $n$.

\item For $p$=1 to $n$, (increasing)

\hspace{1cm} For $j$=$k$ to $1$, (decreasing)
     
\hspace{2cm} $a_j=a_j+X_p~a_{j-1}$.
       
\item The expected value is $\sigma_{k,n}=a_k$.

\end{itemize}

\vspace{0.5cm}

\noindent A transcription in the symbolic algebra language Mathematica \cite{mathematica} reads, for the evaluation of $\sigma_{k,4}$:

\vspace{0.5cm}

\begin{verbatim}
n=4;
a=Table[Boole[i==0],{i,0,n}];
Do[
  a[[j]] = a[[j]] + X[p] a[[j-1]];
  , {p, 1, n},{j, k+1, 2, -1}
]
sigma_kn = a[[k+1]] // Expand   
   
\end{verbatim}

\vspace{0.5cm}

\noindent The number of iterations is $n\times k$. It is also possible (and more compact) to compute directly the recursion relation :

\begin{verbatim}

sigma[k_,n_]:=If[n==0, KroneckerDelta[k,0], sigma[k,n-1] + X[n] sigma[k-1,n-1]]

\end{verbatim}

\vspace{0.5cm}

\item Compute $U_{Q-j,N}\left[\vec{g}-\vec{1}\right]$ using the GP relation (\ref{gp}) \cite{GILLERON04}. 

\end{itemize}

\section{Generalization to the case of average quantities involving $U_Q\left[\vec{g}-\vec{m}_{\alpha}\right]$, $m\ge 2$}\label{sec4}

The case where the degeneracy of a particular orbital $\alpha$ is reduced by more than 1 occurs in the calculation of averages involving powers of subshell populations and deserves special care. 

\begin{itemize}

\item The first possibility consists in using the usual algorithm (GP relation) for $U_Q\left[\vec{g}-\vec{m}_{\alpha}\right]$ without the storage optimisation which is the main purpose of the present work.

\item It is also possible to use relations such as \cite{OREG97}:

\begin{equation}
U_Q\left[\vec{g}-\vec{m}_{\alpha}\right]=\sum_{k=0}^QU_{Q-k}\left[\vec{g}-\vec{m}_{\alpha}+\vec{1}_{\alpha}\right]\left(-X_{\alpha}\right)^k
\end{equation}

\noindent in a recursive way. 

\item Another possibility, following the above procedure relying on the use of the generating function yields, for $m=2$:

\begin{eqnarray}
U_Q\left[\vec{g}-\vec{2}_{\alpha}\right]&=&\left.\frac{1}{Q!}\frac{\partial^Q}{\partial z^Q}\prod_{i=1}^N\left(1+zX_i\right)^{g_i-2\delta_{i\alpha}}\right|_{z=0}\nonumber\\
&=&\left.\frac{1}{Q!}\frac{\partial^Q}{\partial z^Q}\left[\prod_{i=1}^N\left(1+zX_i\right)^{g_i-1}\frac{\prod_{i=1,i\ne\alpha}^N\left(1+zX_i\right)}{1+zX_{\alpha}}\right]\right|_{z=0}\nonumber\\
&=&\left.\frac{1}{Q!}\frac{\partial^Q}{\partial z^Q}\left[F[\vec{g}-\vec{1},z]\frac{H^{(\alpha)}(z)}{1+zX_{\alpha}}\right]\right|_{z=0}.
\end{eqnarray}

\noindent Using the Leibniz rule for the derivative of a product, we get

\begin{eqnarray}
U_Q\left[\vec{g}-\vec{2}_{\alpha}\right]&=&\left.\frac{1}{Q!}\sum_{k=0}^Q\bin{Q}{k}\frac{\partial^{Q-k}}{\partial z^{Q-k}}F[\vec{g}-\vec{1},z].\frac{\partial^k}{\partial z^k}\left[\frac{H^{(\alpha)}(z)}{1+zX_{\alpha}}\right]\right|_{z=0}=\sum_{k=0}^Q\eta_k^{(\alpha)}~U_{Q-k}\left[\vec{g}-\vec{1}\right]\nonumber\\
& &
\end{eqnarray}

\noindent with

\begin{equation}
\eta_k^{(\alpha)}=\left.\frac{1}{k!}\frac{\partial^k}{\partial z^k}\left[\frac{H^{(\alpha)}(z)}{1+zX_{\alpha}}\right]\right|_{z=0}.
\end{equation}

\noindent Applying the Leibniz rule again, we have

\begin{equation}
\eta_k^{(\alpha)}=\left.\frac{1}{k!}\sum_{p=0}^k\bin{k}{p}\frac{\partial^p}{\partial z^p}H^{(\alpha)}(z).\frac{\partial^{k-p}}{\partial z^{k-p}}\left(\frac{1}{1+zX_{\alpha}}\right)\right|_{z=0}
\end{equation}

\noindent and since

\begin{equation}
\left.\frac{\partial^{k-p}}{\partial z^{k-p}}\left(\frac{1}{1+zX_{\alpha}}\right)\right|_{z=0}=(k-p)!\left(-X_{\alpha}\right)^{k-p},
\end{equation}

\noindent we get

\begin{equation}\label{premi}
U_Q\left[\vec{g}-\vec{2}_{\alpha}\right]=\sum_{k=0}^Q\eta_k^{(\alpha)}~U_{Q-k}\left[\vec{g}-\vec{1}\right]
\end{equation}

\noindent with 

\begin{equation}\label{etac}
\eta_k^{(\alpha)}=\sum_{p=0}^kh_p^{(\alpha)}\left(-X_{\alpha}\right)^{k-p},
\end{equation}

\noindent $h_p^{(\alpha)}$ (see Eq. (\ref{hj})) being the elementary symmetric polynomials $\sigma_{p}$. In the Statistical Weight APproximation (SWAP), i.e. $T\rightarrow\infty$ and therefore $X_i\equiv 1$ $\forall i$, we have

\begin{equation}
\bin{G-2}{Q}=\sum_{k=0}^Q\bin{G-N}{Q-k}\sum_{p=0}^k(-1)^{k-p}\bin{N-1}{p}=\sum_{k=0}^Q\bin{G-N}{Q-k}\bin{N-2}{k},
\end{equation}

\noindent which is the Vandermonde identity. This approach can be generalized to $U_Q\left[\vec{g}-\vec{m}_{\alpha}\right]$ with $m\ge 3$, but one may argue that Eq. (\ref{etac}) contains a summation of alternate-sign terms, which may be problematic. In order to avoid that, another approach is possible, which is explained below. 

\item Writing

\begin{eqnarray}
U_Q\left[\vec{g}-\vec{2}_{\alpha}\right]&=&\left.\frac{1}{Q!}\frac{\partial^Q}{\partial z^Q}\prod_{i=1}^N\left(1+zX_i\right)^{g_i-2\delta_{i\alpha}}\right|_{z=0}\nonumber\\
&=&\left.\frac{1}{Q!}\frac{\partial^Q}{\partial z^Q}\left[\prod_{i=1}^N\left(1+zX_i\right)^{g_i-2}\prod_{i=1,i\ne\alpha}^N\left(1+zX_i\right)^2\right]\right|_{z=0}\nonumber\\
&=&\left.\frac{1}{Q!}\frac{\partial^Q}{\partial z^Q}\left\{F[\vec{g}-\vec{2},z]\left[H^{(\alpha)}(z)\right]^2\right\}\right|_{z=0}
\end{eqnarray}

\noindent and using the Leibniz rule, we obtain

\begin{eqnarray}\label{prems}
U_Q\left[\vec{g}-\vec{2}_{\alpha}\right]&=&\left.\frac{1}{Q!}\sum_{k=0}^Q\bin{Q}{k}\frac{\partial^{Q-k}}{\partial z^{Q-k}}F[\vec{g}-\vec{2},z].\frac{\partial^k}{\partial z^k}\left[H^{(\alpha)}(z)\right]^2\right|_{z=0}=\sum_{k=0}^Q\epsilon_k^{(\alpha)}[2]\times U_{Q-k}\left[\vec{g}-\vec{2}\right],\nonumber\\
\end{eqnarray}

\noindent with

\begin{equation}
\epsilon_k^{(\alpha)}[2]=\left.\frac{1}{k!}\frac{\partial^k}{\partial z^k}\left[H^{(\alpha)}(z)\right]^2\right|_{z=0}.
\end{equation}

\noindent The determination of $\epsilon_k^{(\alpha)}[2]$ is detailed in \ref{appc}. We have (like for quantities $h_j^{(\alpha)}$ and $\sigma_{i,n}$ above, we intentionally omit $\alpha$ and add the dependence with respect to the number of subshells $n$):

\begin{equation}\label{epsilon}
\epsilon_{j,n}[2]=\epsilon_{j,n-1}[2]+2X_n~\epsilon_{j-1,n-1}[2]+X_n^2~\epsilon_{j-2,n-1}[2]
\end{equation}

\noindent initialized with $\epsilon_{j,0}[2]=1$. Equations (\ref{prems}) and (\ref{epsilon}) can be easily generalized to other values of $m$. For instance, for $m$=3, one has

\begin{equation}\label{premi}
U_Q\left[\vec{g}-\vec{3}_{\alpha}\right]=\sum_{k=0}^Q\epsilon_k^{(\alpha)}[3]\times U_Q\left[\vec{g}-\vec{3}\right],
\end{equation}

\noindent with

\begin{equation}\label{eta}
\epsilon_{j,n}^{(\alpha)}[3]=\epsilon_{j,n-1}^{(\alpha)}[3]+3X_n~\epsilon_{j-1,n-1}^{(\alpha)}[3]+3X_n^2~\epsilon_{j-2,n-1}^{(\alpha)}[3]+\epsilon_{j-3,n-1}^{(\alpha)}[3]
\end{equation}

\noindent and for any value of $m$:

\begin{equation}\label{premm}
U_Q\left[\vec{g}-\vec{m}_{\alpha}\right]=\sum_{k=0}^Q\epsilon_k^{(\alpha)}[m]\times U_{Q-k}\left[\vec{g}-\vec{m}\right],
\end{equation}

\noindent with (here also we intentionally omit $\alpha$ and add the dependence with respect to the number of subshells $n$):

\begin{equation}
\epsilon_{j,n}[m]=\sum_{p=0}^m\bin{m}{p}X_n^p~\epsilon_{j-p,n-1}[m].
\end{equation}

\noindent In the SWAP approximation, Eq. (\ref{premm}) becomes

\begin{equation}
\bin{G-m}{Q}=\sum_{k=0}^ Q\bin{G-mN}{Q-k}\bin{m(N-1)}{k},
\end{equation}

\noindent which is also the Vandermonde identity. It is worth mentioning that coefficients $\epsilon_{j,n}[m]$ are related to symmetric polynomials. As mentioned above, $\epsilon_{j,n}[1]=h_k^{(\alpha)}$ are elementary symmetric polynomials $\sigma_{j,n}=\mathcal{P}_j(X_1,X_2, \cdots, X_n)$ where $\mathcal{P}_j$ is the $j^{th}$ general symmetric polynomial, but actually one has also the relations:

\begin{equation}
\begin{array}{l}
\epsilon_{j,n}[2]=\mathcal{P}_j(X_1,X_1,X_2,X_2, \cdots, X_n,X_n)\\
\epsilon_{j,n}[3]=\mathcal{P}_j(X_1,X_1,X_1,X_2,X_2,X_2, \cdots, X_n,X_n,X_n)\\
\epsilon_{j,n}[4]=\mathcal{P}_j(X_1,X_1,X_1,X_1,X_2,X_2,X_2,X_2, \cdots, X_n,X_n,X_n,X_n)\\
\vdots\\
\epsilon_{j,n}[m]=\mathcal{P}_j(\underbrace{X_1, \cdots, X_1}_{\mathrm{m~ times}},\underbrace{X_2, \cdots, X_2}_{\mathrm{m~ times}}, \cdots, \underbrace{X_n, \cdots, X_n}_{\mathrm{m~ times}}).\\
\end{array}
\end{equation}

\end{itemize}

\section{Conclusion}

An efficient extension of our stable recursion relation for the determination of partition functions of a canonical ensemble of non-interacting bound electrons was published three years after the original work. However, the procedure described in the latter article was not easy to implement in practise. This is mainly due to the fact that Eq. (23) of Ref. \cite{WILSON07} was written in a very compact way, which makes it difficult to understand. In this short paper, we provided details of the derivation of the main equations and gave an example, in order to facilitate the practical use. We also pointed out the fact that the algorithm involves elementary symmetric polynomials. That was not mentioned in Ref. \cite{WILSON07}, and makes possible the derivation of further new relations. The method enables one to calculate efficiently averages such as

\begin{equation}
\langle n_{\alpha}^an_{\beta}^bn_{\gamma}^c\cdots\rangle,
\end{equation}

\noindent which are the cornerstone of the statistical modeling of complex absorption and emission spectra in the framework of the superconfiguration approach. All the relations discussed in the present paper can be modified by effective statistical weights in order to account for pressure ionization \cite{BUSQUET13}. They can also be applied together with the inclusion of electron-electron interactions using the Jensen-Feynman approach, following Ref. \cite{PAIN09}.

\appendix

\section{Recurrence relation for coefficients $\sigma_{j,n}$}\label{appa}

Equation (\ref{hj}) becomes (we replace, for the sake of simplicity, $h$ by $\sigma$, which is the usual notation for elementary symmetric polynomials):

\begin{equation}
\sigma_{j,n}=\lim_{z\rightarrow 0}\frac{1}{j!}\frac{\partial^j}{\partial z^j}H_n(z),
\end{equation}

\noindent where the generating function $H_n(z)=\prod_{i=1}^n\left(1+zX_i\right)$ is in fact a particular case of the generating function of the partition functions $U_Q$ \cite{GILLERON04,WILSON07,PAIN11}, which is

\begin{equation}
\prod_{i=1}^n\left(1+zX_i\right)^{g_i},
\end{equation}

\noindent but with $g_i$=1 $\forall i$. Therefore the BS relation for partition functions

\begin{equation}
U_{Q,N}[\vec{g}]=\frac{1}{Q}\sum_{k=0}^Q(-1)^{k+1}~\chi_{k,N}~U_{Q-k,N}[\vec{g}],
\end{equation}

\noindent with $\chi_{k,N}=\sum_{s=1}^Ng_kX_s^k$ becomes 

\begin{equation}\label{bslike}
\sigma_{j,n}=\frac{1}{j}\sum_{k=1}^j(-1)^{k+1}~\chi_{k,n}~\sigma_{j-k,n},
\end{equation}

\noindent where $\chi_{k,n}=\sum_{s=1}^nX_s^k$ and $\sigma_{0,n}=1$. Equation (\ref{bslike}) is known as the Newton-Girard identity \cite{SEROUL00}. The problem is that the latter relation contains alternate sums, which can lead to numerical instabilities \cite{GILLERON04}, especially at low temperature. Using the expression of the multiple derivative of a product of functions, one has

\begin{eqnarray}\label{mult}
\sigma_{j,n}=\lim_{z\rightarrow 0}\frac{1}{j!}\sum_{\tiny\begin{array}{c}
i_1, i_2, \cdots, i_n=0\\
i_1+i_2+\cdots +i_n=j
\end{array}}^1\bin{j}{i_1i_2\cdots i_n}\frac{\partial}{\partial z^{i_1}}\left(1+zX_1\right).\frac{\partial}{\partial z^{i_2}}\left(1+zX_2\right)\cdots\frac{\partial}{\partial z^{i_n}}\left(1+zX_n\right),\nonumber\\
& &
\end{eqnarray}

\noindent where

\begin{equation}
\bin{j}{i_1i_2\cdots i_n}=\frac{j!}{i_1!i_2!\cdots i_n!}
\end{equation}

\noindent is the multinomial coefficient. Therefore, one has

\begin{equation}\label{mult2}
\sigma_{j,n}=\frac{1}{j!}\sum_{\tiny\begin{array}{c}
i_1, i_2, \cdots, i_n=0\\
i_1+i_2+\cdots +i_n=j
\end{array}}^1\bin{j}{i_1i_2\cdots i_n}\prod_{s=1}^n\left[1+(X_s-1)i_s\right].
\end{equation}

\noindent Since Eq. (\ref{mult2}) can be rewritten as

\begin{equation}
\sigma_{j,n}=\frac{1}{j!}\sum_{i_n=0}^1\frac{\left[1+(X_n-1)i_n\right]}{i_n!}\sum_{\tiny\begin{array}{c}
i_1, i_2, \cdots, i_{n-1}=0\\
i_1+i_2+\cdots i_{n-1}=j-i_n
\end{array}}^1\bin{j}{i_1i_2\cdots i_{n-1}}\prod_{s=1}^{n-1}\left[1+(X_s-1)i_s\right],
\end{equation}

\noindent we get the following recursion relation

\begin{equation}
\sigma_{j,n}=\sigma_{j,n-1}+X_n\sigma_{j-1,n-1}.
\end{equation}

\section{Additional considerations about the link between partition functions and symmetric polynomials}\label{appb}

Writing

\begin{equation}
H_{n-1}(z)=\frac{\prod_{i=1}^n\left(1+zX_i\right)}{\left(1+zX_n\right)}=\frac{H_n(z)}{\left(1+zX_n\right)}\
\end{equation}

\noindent and applying the Leibniz multiple derivation formula of the product of two functions, we get

\begin{equation}
\frac{\partial^jH_{n-1}(z)}{\partial z^j}=\sum_{k=0}^j\bin{j}{k}\frac{\partial^k}{\partial z^k}\frac{1}{\left(1+zX_n\right)}.\frac{\partial^{j-k}H_{n}(z)}{\partial z^{j-k}}
\end{equation}

\noindent yielding

\begin{equation}\label{oreg}
\sigma_{j,n-1}=\sum_{k=0}^j\left(-X_n\right)^k\sigma_{j-k,n},
\end{equation}

\noindent which is a particular case of

\begin{equation}
U_{Q,N-1}=\sum_{k=0}^Q\bin{g_N+k-1}{g_N-1}\left(-X_N\right)^kU_{Q-k,N},
\end{equation}

\noindent with $g_i$=1 $\forall i$ in the case of partition functions. However, Eq. (\ref{oreg}) involves alternate sums, which can be responsible for numerical instabilities, as for relation (\ref{bslike}). Therefore, the most efficient recursion relation is (\ref{recuelem}). The method can be further optimized, since there is an abundant literature about the fast computation of elementary symmetric polynomials (see for instance Refs. \cite{LAUER76,ZEILBERGER85,MEAD92,MINAS03,RESPONDEK13,JIANG13}). In particular, it is worth pointing out that the BS relation in the case where all the degeneracies are equal to one (see Eq. (\ref{bslike}) in Appendix A), is nothing else than the so-called Newton-Girard identity \cite{SEROUL00}, relating the elementary symmetric polynomials to the power sums $\chi_{k,n}$ defined as

\begin{equation}
\chi_{k,n}=\sum_{s=1}^nX_s^k.
\end{equation}

\noindent In fact, one can write

\begin{equation}\label{sym}
\sigma_{j,n}=(-1)^j\sum_{\tiny
\begin{array}{c}
m_1, m_2,\cdots, m_n\geq 0\\ 
m_1+2m_2+\cdots +jm_j=j
\end{array}
}\prod_{k=1}^j\frac{\left(-\chi_{k,n}\right)^{m_k}}{m_k!k^{m_k}}
\end{equation}

\noindent or using the Bell polynomials $B_j$:

\begin{equation}\label{bell}
\sigma_{j,n}=\frac{(-1)^j}{j!}B_j\left(-\chi_{1,n},-1!\chi_{2,n},-2!\chi_{3,n}, \cdots,-(j-1)!\chi_{j,n}\right).
\end{equation}

\noindent The elementary symmetric polynomials can also be expressed as determinants:

\begin{equation}\label{det}
\sigma_{j,n}=\frac{1}{j!}\left|
\begin{array}{ccccc}
\chi_{1,n} & 1 & 0 & \cdots & \\
\chi_{2,n} & \chi_{1,n} & 2 & 0 & \cdots \\
\cdots & \cdots & \ddots & \ddots & \vdots \\
\chi_{j-1,n} & \chi_{j-2,n} & \cdots & \chi_{1,n} & j-1 \\
\chi_{j,n} & \chi_{j-1,n} & \cdots & \chi_{2,n} & \chi_{1,n} \\
\end{array}
\right|.
\end{equation}

\noindent Relations such as (\ref{bell}) and (\ref{det}) may give new ideas of additional relations and algorithms. 

\section{Recurrence relation for coefficients $\epsilon_{j,n}[2]$}\label{appc}

We have

\begin{equation}
\epsilon_{j,n}[2]=\lim_{z\rightarrow 0}\frac{1}{j!}\frac{\partial^j}{\partial z^j}\left[H_n(z)\right]^2,
\end{equation}

\noindent i.e.

\begin{eqnarray}
\epsilon_{j,n}[2]=\lim_{z\rightarrow 0}\frac{1}{j!}\sum_{\tiny\begin{array}{c}
i_1, i_2, \cdots, i_n=0\\
i_1+i_2+\cdots +i_n=j
\end{array}}^2\bin{j}{i_1i_2\cdots i_n}\frac{\partial}{\partial z^{i_1}}\left(1+zX_1\right)^2.\frac{\partial}{\partial z^{i_2}}\left(1+zX_2\right)^2\cdots\frac{\partial}{\partial z^{i_n}}\left(1+zX_n\right)^2,\nonumber\\
& &
\end{eqnarray}

\noindent yielding

\begin{equation}\label{mult3}
\epsilon_{j,n}[2]=\frac{1}{j!}\sum_{\tiny\begin{array}{c}
i_1, i_2, \cdots, i_n=0\\
i_1+i_2+\cdots +i_n=j
\end{array}}^2\bin{j}{i_1i_2\cdots i_n}\prod_{s=1}^n\mathcal{D}_s,
\end{equation}

\noindent with $\mathcal{D}_s=(i_s-1)(i_s-2)/2-2i_s(i_s-2)X_s+i_s(i_s-1)X_s^2$. Since Eq. (\ref{mult3}) can be rewritten as

\begin{equation}
\epsilon_{j,n}[2]=\frac{1}{j!}\sum_{i_n=0}^2\frac{\mathcal{D}_n}{i_n!}\sum_{\tiny\begin{array}{c}
i_1, i_2, \cdots, i_{n-1}=0\\
i_1+i_2+\cdots i_{n-1}=j-i_n
\end{array}}^2\bin{j}{i_1i_2\cdots i_{n-1}}\prod_{s=1}^{n-1}\mathcal{D}_s,
\end{equation}

\noindent  we get the following recursion relation

\begin{equation}
\epsilon_{j,n}[2]=\epsilon_{j,n-1}[2]+2X_n~\epsilon_{j-1,n-1}[2]+X_n^2~\epsilon_{j-2,n-1}[2].
\end{equation}

\section*{Acknowledgments}

The authors would like to thank M. Busquet for helpful discussions about the practical implementation of the method and Y. Kurzweil for useful criticism and suggestions of improvements.

\end{document}